\documentclass[useAMS]{mn2e}
\usepackage{rotating}
\usepackage{mncite}

\newcommand\msun {M$_{\odot}$}
\def\approxgt{\ifmmode \rlap{$>$}{}_{{}_{{}_{\textstyle\sim}}} \else%
$\rlap{$>$}{}_{{}_{{}_{\textstyle\sim}}}$\fi} 
\def\approxlt{\ifmmode \rlap{$<$}{}_{{}_{{}_{\textstyle\sim}}} \else%
$\rlap{$<$}{}_{{}_{{}_{\textstyle\sim}}}$\fi}

\LARGE 
\normalsize 
\title[A search for the counterpart of the accreting millisecond X--ray pulsar XTE~J1751--305]{A search for the optical and near--infrared counterpart of the accreting millisecond X--ray pulsar XTE~J1751--305}

\author[P.G. Jonker et al.]  {P.G. Jonker$^1$\thanks{email :
    peterj@ast.cam.ac.uk; based on observations collected at the
    European Southern Observatory, Chile (ESO No 269.D-5013)}, G.
  Nelemans$^1$, Z. Wang$^2$, A.K.H. Kong$^3$, D.  Chakrabarty$^2$, M.
  Garcia$^3$, \newauthor P.J. Groot$^4$, M. van der Klis$^5$, T.
  Kerr$^9$, B. Mobasher$^{10}$, M. Sullivan$^{11}$, T. Augusteijn$^6$,
  \newauthor B.W. Stappers$^{5,12}$, P. Challis$^3$, R.P. Kirshner$^3$, J. Hjorth$^7$, A. Delsanti$^8$ \\
  $^1$Institute of Astronomy, Madingley Road, CB3 0HA, Cambridge, UK\\
  $^2$Department of Physics and Center for Space Research, Massachusetts Institute of Technology, Cambridge, MA 02139, USA\\
  $^3$Harvard--Smithsonian Center for Astrophysics, 60 Garden
  Street, Cambridge, MA 02138, USA\\
  $^4$Department of Astrophysics, University of Nijmegen, P.O.Box
  9010,
  Nijmegen, The Netherlands\\
  $^5$Astronomical Institute ``Anton Pannekoek'',
  University of Amsterdam, Kruislaan 403, 1098 SJ Amsterdam, The Netherlands\\
  $^6$Nordic Optical Telescope, Apartado 474, 38700 Santa Cruz de La Palma, Canary Islands, Spain \\
  $^7$Astronomical Observatory, University of Copenhagen, Juliane Maries Vej 30, DK-2100 Copenhagen {\O}, Denmark\\
  $^8$European Southern Observatory, Alonso de Cordova 3107, Vitacura, Casilla 19001, Santiago 19, CHILE\\
  $^9$Joint Astronomy Centre, University Park, 660 N. A'ohoku Place, Hilo, Hawaii 96720, USA\\
  $^{10}$Space Telescope Science Institute, 3700 San Martin Drive, Baltimore MD 21218, USA \\
  $^{11}$Physics Department University of Durham, Science Labs, South
  Road, Durham, DH1 3LE, UK\\
  $^{12}$Dwingeloo, Postbus 2, 7990 AA, Dwingeloo, The Netherlands}

\begin{document}

\maketitle

\begin{abstract}
\noindent 
We have obtained optical and near--infrared images of the field of the
accreting millisecond X--ray pulsar XTE~J1751--305. There are no stars
in the 0.7'' error circle (0.7'' is the overall uncertainty arising
from tying the optical and X--ray images and from the intrinsic
uncertainty in the {\it Chandra} X--ray astrometric solution). We
derive limiting magnitudes for the counterpart of R$>$23.1, I$>$21.6,
Z$>$20.6, J$>$19.6, K$>$19.2. We compare these upper limits with the
magnitudes one would expect for simple models for the possible donor
stars and the accretion disk subject to the reddening observed in
X--rays for XTE~J1751--305 and when put at the distance of the
Galactic Centre (8.5 kpc). We conclude that our non--detection does
not constrain any of the models for the accretion disk or possible
donor stars. Deep, near--infrared images obtained during quiescence
will, however, constrain possible models for the donor stars in this
ultra--compact system.

\end{abstract}


\section{Introduction}
\label{intro}
Low--mass X--ray binaries (LMXBs) are typically old ($>10^8$ yr)
binary systems in which a low--mass companion star ($\approxlt
1\,M_{\odot}$) transfers matter to a neutron star or a black hole. The
neutron star LMXBs are thought to be among the predecessors of the
millisecond radio pulsars. Due to accretion of matter and decay of the
magnetic field during the LMXB phase, the neutron star spins--up to
millisecond periods (see \pcite{bhatta1995} for a review). However,
until March 2002 only six LMXBs were known to show pulsations (see
\pcite{2001ApJ...553L..43J} for a short overview), and only one of
them, a transient system, was shown to have a millisecond period
(SAX~J1808.4--3658; \pcite{1998Natur.394..344W} and
\pcite{1998Natur.394..346C}). In 2002 \scite{2002IAUC.7867....1M} (see
also \pcite{2002ApJ...575L..21M}) and \scite{2002IAUC.7900....2G} (see
also \pcite{2002ApJ...576L.137G}) discovered millisecond pulsations in
two other transient sources, XTE~J1751--305 and XTE~J0929--314, in
outburst.

There are two main reasons to search for the companion stars of
accretion powered millisecond X--ray pulsars. Accretion powered
millisecond X--ray pulsars have most likely accreted a substantial
amount of matter as they were spun--up by accretion to a period of
milliseconds. A spectroscopical determination of the radial velocity
curve of the companion star yields a lower limit to the mass of the
neutron star; measuring a mass of considerably more than 1.4 M$_\odot$
for even one neutron star would imply the firm rejection of many
proposed equations of state (see the discussion by
\pcite{1995xrb..book...58V}; for an overview of the mass
determinations of millisecond radio pulsars in neutron star--white
dwarf binaries see \pcite{1999ApJ...512..288T}). In order to
spectroscopically measure the radial velocity of the companion star,
clearly a detection of the object is first required.  Secondly, at
present there are three known ways to produce compact X--ray binaries
(\pcite{1986AA...155...51S,1989SvA....33..606T,2002A&A...388..546Y,1987SvAL...13..328T,2002ApJ...565.1107P};
see Section 3 for a more elaborate account of these models).
Detection of the companion star will allow us to differentiate between
these different formation scenarios.

In this Paper we present the results of our search for the optical and
near--infrared counterpart of the ultra--compact accreting millisecond
X--ray pulsar XTE~J1751--305 (${\rm P_{orb}}=42$ minutes;
\pcite{2002ApJ...575L..21M}).  Early reports on the data presented in
this Paper were published by \scite{2002IAUC.7872....2J},
\scite{2002IAUC.7874....2W}, \scite{2002IAUC.7880....2W}, and
\scite{2002ATel...87....1K}.

\section{Observations and analysis}

We observed the region around the accurate {\it Chandra} X--ray
position of the source using the 6.5~m Magellan, the 3.8~m UKIRT, the
3.58~m NTT, and the 1.54~m Danish telescopes. A spectrum of the
candidate optical counterpart presented by \scite{2002ATel...87....1K}
was obtained with ESO's 3.6~m telescope at La Silla using Grism number
12 which covers 601--1032 nm. A log of the observations can be found
in Table~\ref{log}. In case of the near--infrared data the total
exposure time is given {\it and} the dither pattern, i.e. in case of
the UKIRT J band images we took 5 exposures of 60 seconds each, in
between each observation the telescope was dithered; this pattern was
repeated twice. In the case of the Magellan near--infrared
observations we obtained J and Ks images of 60 seconds integration
time each. We also obtained 11 J band images of 20 seconds integration
each.

The data reduction was performed in \textsc{iraf}\footnote{\textsc
  {iraf} is distributed by the National Optical Astronomy
  Observatories}. The near--infrared images (J and K band) were sky
subtracted (using the sky determined from the dithered images),
flatfielded, aligned, and combined to form one image per band per
night. The Magellan near--infrared data have been linearised before
the reduction. The optical images (R, I, and Z band) were reduced in
the standard fashion. Aperture and point spread function fitting
photometry were done using the packages \textsc{apphot} {\rm and}
\textsc{daophot} in \textsc{iraf}.  The spectrum obtained with ESO's
3.6~m telescope was extracted using the task \textsc{apall} in the
package \textsc{specred} in \textsc{iraf}. All nights were photometric
except for the nights of April 13 and 14, 2002 when the humidity was
very high at Magellan and April 14, 2002 at UKIRT when thin cirrus was
present. The J band observations during that night also suffered from
a bright sky as they were obtained partially during twilight. Charge
overflow from a neighbouring bright star covered the location of
XTE~J1751--305 on the CCD in our image obtained with the Danish
telescope, rendering it impossible to determine whether a source is
present or not at the {\it Chandra} X--ray location. Hence, we will
not consider this observation any further.

\begin{table*}
\caption{Log of the observations. MJD and start time refer to the time at the start of the first observation.}
\label{log}
\begin{center}

\begin{tabular}{lccccccc}
\hline
Telescope & Instrument & Observation date  & MJD  & Filters & Exposure & Airmass & Seeing \\ 
  &              & and start time (UT)  & (UTC) & Bessel (except J, K(s))  & Time (s) & & (arcseconds) \\
\hline
\hline
Danish   & DFOSC & 08--04--2002 08:51 & 52372.36870 & R & 900 & 1.01 & $\sim$1.2\\
Magellan & Classic--CAM & 09--04--2002 08:02 & 52373.33472 & Ks & 60  & 1.05&$\sim$0.65\\
Magellan & Classic--CAM & 09--04--2002 07:32 & 52373.31388 & J & 280 (11x20+60) & 1.05 & $\sim$0.65\\
Magellan & MagIC & 13--04--2002 07:43 & 52377.32153 & I & 240 (3x) & 1.05 & $\sim$0.6\\
Magellan & MagIC & 14--04--2002 08:30 & 52378.35416 & I & 30 (1x), 240 (18x) & 1.00--1.03 & $\sim$0.7\\
UKIRT    & UFTI   & 14--04--2002 14:41 & 52378.61201 & J & 600 (2x5x60) & 1.6 & $\sim$0.8\\
UKIRT    & UFTI   & 14--04--2002 14:55 & 52378.62161 & K & 600 (2x5x60) & 1.6 & $\sim$0.65\\
UKIRT    & UFTI   & 18--04--2002 14:47 & 52382.61623 & J & 600 (2x5x60) & 1.6 & $\sim$0.65\\
UKIRT    & UFTI   & 18--04--2002 15:06 & 52382.62925 & K & 600 (2x5x60) & 1.6 & $\sim$0.5\\
NTT      & SUSI2   & 18--04--2002 08:33 & 52382.35682 & R & 600 & 1.00 & $\sim$0.8\\
NTT      & SUSI2   & 18--04--2002 08:23 & 52382.34950 & I & 600 & 1.00 & $\sim$0.9\\
NTT      & SUSI2   & 18--04--2002 08:12 & 52382.34185 & Z & 600 & 1.01 & $\sim$0.8\\
3.6M (spec)& EFOSC  & 01--05--2002 07:19 & 52395.30470 & Gr\#12 0.7'' slit & 1500 & 1.01 & $\sim$0.9\\
\end{tabular}
\end{center}

\end{table*}

We derived an astrometric solution for the optical I band image
obtained with the NTT of the field of XTE~J1751--305 using the
positions of 4 nearby unsaturated stars which appear in the USNO-A1.0
catalogue. The rms of the fit was 0.015''. The typical astrometric
error of stars in the USNO--A1.0 catalogue is 0.25'' (68 per cent
confidence); the uncertainty in the X--ray position is dominated by
the {\it Chandra} aspect solution (0.6''; 90 per cent confidence;
\pcite{2002ApJ...575L..21M}).  Hence, the overall astrometric
uncertainty in the position of the errorcircle is 0.7'' (90 per cent
confidence).  Next, we tied the astrometric solution of the I band
image to the optical R and Z band and to the near--infrared J and K(s)
band images by assigning the known position of several stars (from the
I band) to the R, Z, J(s), or K band image; this did not increase the
error in the astrometry significantly.

\begin{figure*}
  \includegraphics[width=14cm]{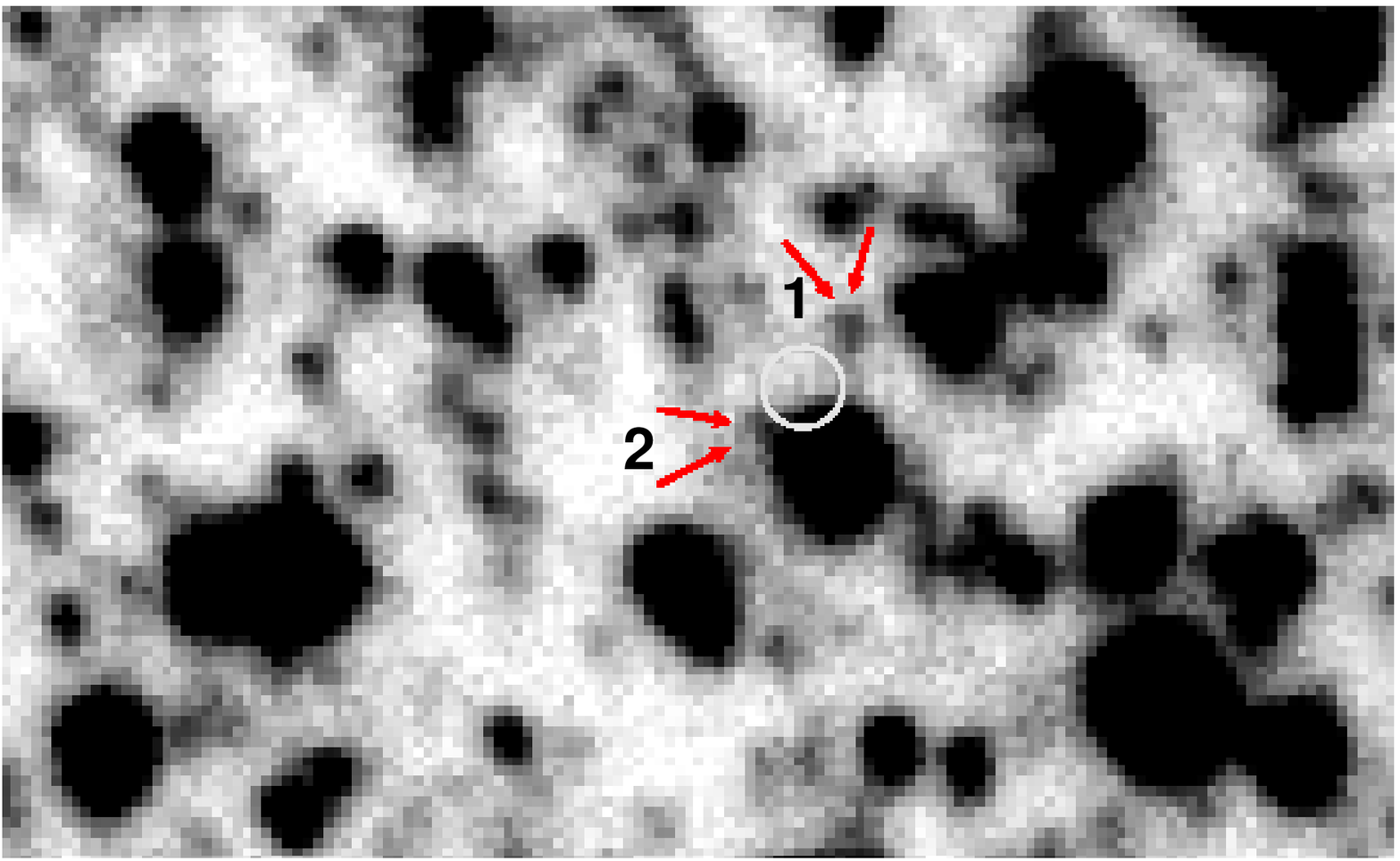}
\quad
\includegraphics[width=14cm]{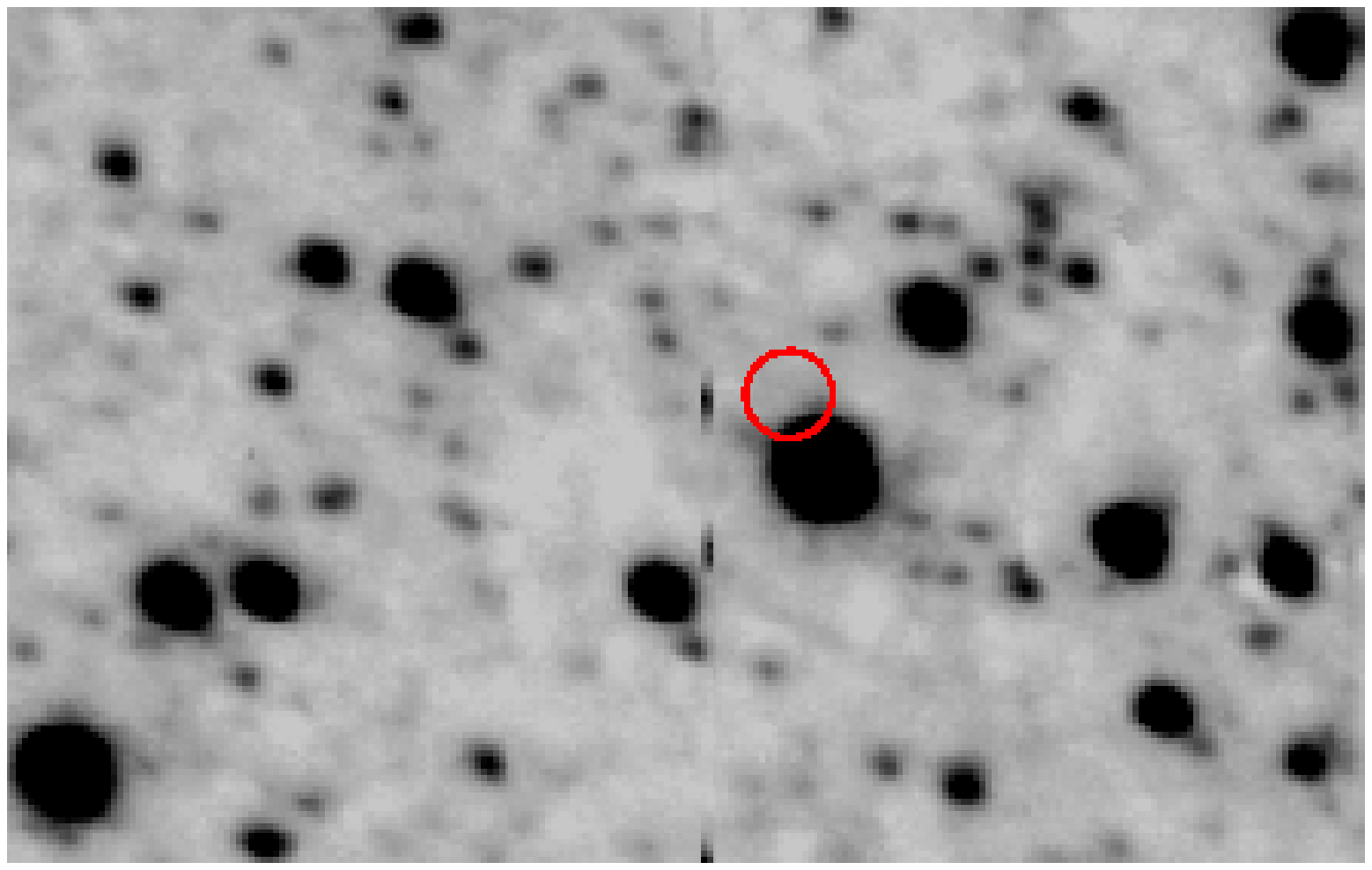}
\caption{{\it Top panel:} The R band image ($\sim 27''\times 14''$; 10 minutes
  integration; North is up, East to the left) of the region of
  XTE~J1751--305 obtained with the 3.58~m NTT. The X--ray error circle
  for the location of the source is overplotted (0.7''; 90 per cent
  confidence). The double set of arrows indicate the location of the
  possible counterparts proposed by Wang \& Chakrabarty (2002; star 1)
  and Kong et al.  (2002; star 2). {\it Bottom panel:} The UKIRT K
  band image obtained on April 18, 2002 ($\sim 27''\times 14''$; 10
  minutes integration; North is up, East is to the left) of the region
  of XTE~J1751--305. The {\it Chandra} error circle is overplotted.
  The vertical black strips in the centre of the image are artifacts
  introduced when combining the dithered images.}
\label{images}
\end{figure*}

In Figure~\ref{images} ({\it top panel}) we show the R band image from
our NTT observations. Separately, in Figure~\ref{images} ({\it bottom
  panel}) we show the near--infrared UKIRT (K band) image. From the
astrometry it is clear that each of the two candidate counterparts
(both indicated with two arrows in the {\it top panel}) is just
outside the 90 per cent confidence error circle
(\pcite{2002IAUC.7874....2W} initially proposed star 1 as the
counterpart but see also \pcite{2002IAUC.7880....2W};
\pcite{2002ATel...87....1K} proposed star 2 as the counterpart; see
Figure~\ref{images}).  Hence, it is unlikely but not ruled out that
the counterpart was detected.

We determined 3 $\sigma$ upper limits on the presence of a star in the
error circle of each of the images obtained under photometric
conditions. To do this we added a simulated artificial star (created
using the point spread function of the stars in the image) at the
position of the error circle. The magnitude of this extra star was
varied and measured using the standard photometric tasks. We define
the 3 $\sigma$ limiting magnitude as the magnitude at which the error
in the magnitude of such an artificial star is $\sim$0.3 magnitudes
(an 0.3 magnitude error is equivalent to a $\sim$30 per cent error on
the flux measurement, i.e. a three sigma detection; see also
\pcite{2000A&A...358..605H}). For all the images we need $\sim$3--5
trials to cover the error circle (the number of trials is different
for the different images; it varies as a function of the seeing).  We
did not take the number of trials into account when calculating the
upper limits (the error introduced by this is $\sim$0.1 magnitude).
The error on the determined limiting magnitude is also $\sim$0.1
magnitude. Together with the uncertainty in the photometric
zero--point (at most 0.1 magnitude) this yields an uncertainty of
$\sim$0.2 magnitude in the limiting magnitudes.

For the NTT Z band image we converted the (Sloan) z' magnitude of the
observed standard star (PG0918+029D) to the UKIRT UFTI Z band using
the transformation given on the UKIRT web page. However, since the NTT
and the UKIRT Z band differ, and there is no cross-calibration of
these two Z bands, in doing so we introduced an uncertainty in the Z
band magnitudes which can be more than 0.2 magnitudes.  Therefore, the
Z band upper limit is given as reference only; it should be considered
approximate. The upper limits are given in Table~\ref{upper}.

\begin{table*}
\caption{Upper limits (3 $\sigma$) on the presence of a star at the 
position of the {\it Chandra} error circle.}
\label{upper}
\begin{center}

\begin{tabular}{lcc}
\hline
Telescope & Observation date  & Limiting magnitude \\ 
\hline
\hline
Magellan & April, 9, 2002 & J $>$19.6 \\
Magellan & April, 9, 2002 & K $>$18.2 \\
UKIRT & April 18, 2002 & J $>$19.5 \\
UKIRT & April 18, 2002 & K $>$19.2 \\
NTT &  April 18, 2002 & R$>$23.1 \\
NTT &  April 18, 2002 & I$>$21.6 \\
NTT &  April 18, 2002 & Z$>$20.6 \\
\end{tabular}
\end{center}
\end{table*}

We searched for variability in the I band magnitude for the candidate
counterpart proposed by \scite{2002IAUC.7880....2W} (star 1) by
comparing the magnitudes in the Magellan and the NTT observations.
The I band magnitude of star 1 is consistent with being the same
during the observations. Unfortunately, the I band magnitudes of the
candidate counterpart proposed by \scite{2002ATel...87....1K} (star 2)
could not be determined for the NTT image due to the presence of the
nearby star. However, differential photometry showed that star 2 was
not variable in the Magellan Sloan i' band images taken a day apart
(\pcite{2002ATel...87....1K}).

\begin{figure*}
  \includegraphics[width=17cm,height=12cm]{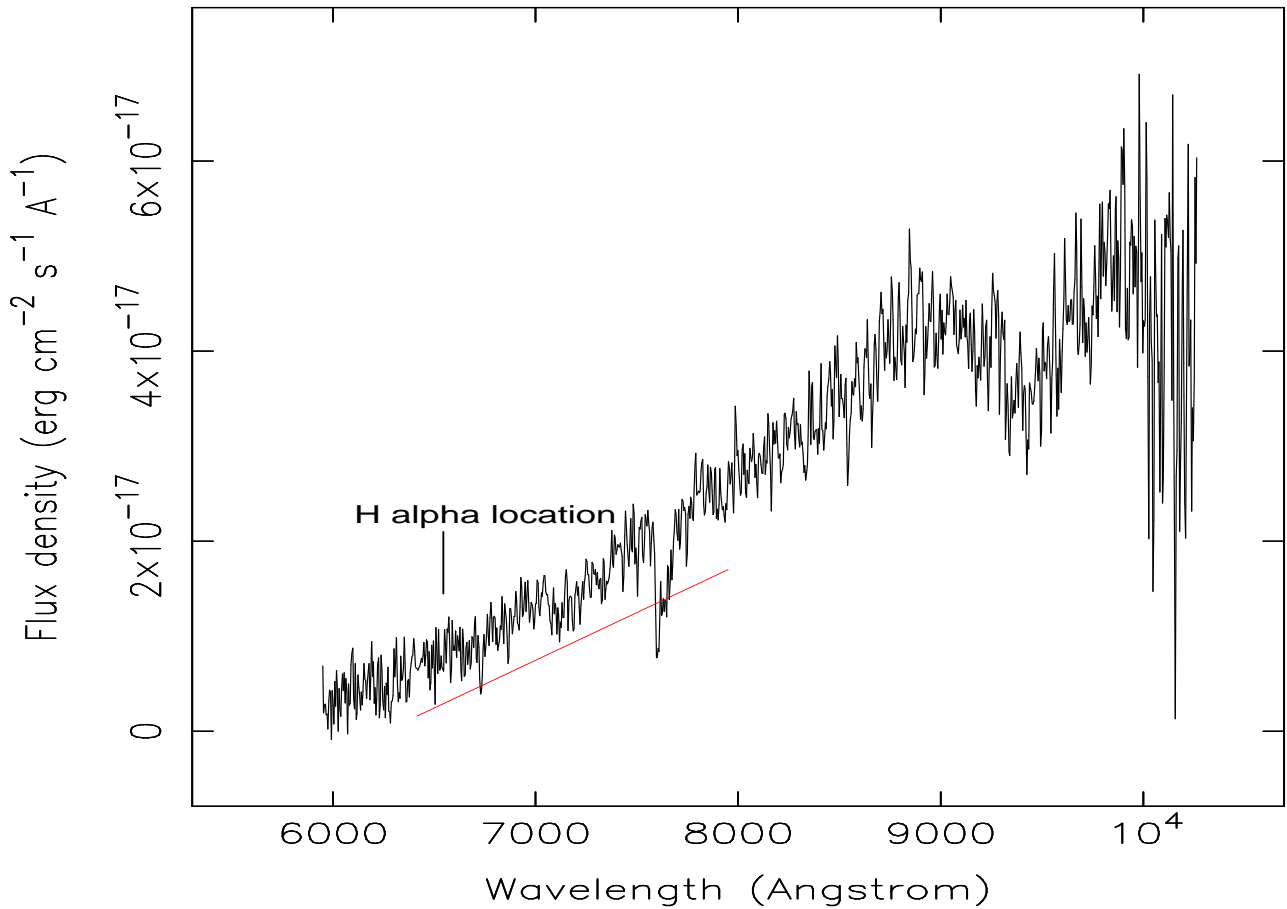}
\caption{Spectrum of the Southernmost source identified with the marks in 
  Figure~\ref{images} (star 2; to the East of the bright neighbour),
  obtained with ESO's 3.6~m telescope at La Silla on May 1, 2002. The
  line connects the magnitudes of this star in the R and I band
  converted to flux densities at the central wavelength of the R and I
  band filter.}
\label{spec}
\end{figure*}

The slit we used to obtain a spectrum using ESO's 3.6~m telescope at
La Silla had a width of 0.7''.  The slit orientation was such that
both star 1 and 2 were in the slit as well as most of the position
marked by the error circle.  However, due to the seeing of $\sim$0.9''
during these observations the spectrum of star 2 may have been
contaminated by light of the bright nearby star. In order to
investigate this we converted the magnitudes of this bright nearby
star (R=18.3, I=16.2) and those of star 2 (R=22.6, I=19.2; we note
that the uncertainty on these magnitudes is large due to the
non--photometric conditions at the time of the Magellan I band
observations; for comparison the best estimate of the magnitude of
star 1 was 20.2) to flux densities and we compared the amplitude of
the spectral energy distribution according to these R and I band flux
densities with the amplitude of the flux calibrated spectrum. The fact
that the flux density of the spectrum is higher than that of star 2
shows that the spectrum will have been contaminated with light from
the bright nearby star. We note that in converting the magnitudes to
fluxes we assumed that star 2 was not variable. We show the spectrum
in Figure~\ref{spec}. We also label the position of H$\alpha$,
although it is unclear whether the donor star of XTE~J1751--305
contains hydrogen or not. The spectrum is featureless except for the
atmospheric absorption feature near 7613 Angstrom and the band at
$\sim$9300 Angstrom. The spectrum of star 1 was not detected,
rendering further support to the conclusion that the detected light
from the position of star 2 was dominated by the bright nearby star.

\section{Discussion}

We have obtained optical and near--infrared images of the field of the
accreting millisecond X--ray pulsar XTE~J1751--305. Two possible
counterparts have been investigated but since they both fall outside
the 90 per cent confidence {\it Chandra} error circle we conclude that
neither the optical nor the near--infrared counterpart was detected.
The star closest to the error circle was suggested to be the
counterpart by \scite{2002ATel...87....1K} (star 2 in
Figure~\ref{images}). We placed upper limits on the presence of a star
in the error circle in the R, I, Z, J, and K band.

\begin{figure}
  \resizebox{\columnwidth}{!}{\includegraphics[angle=-90]{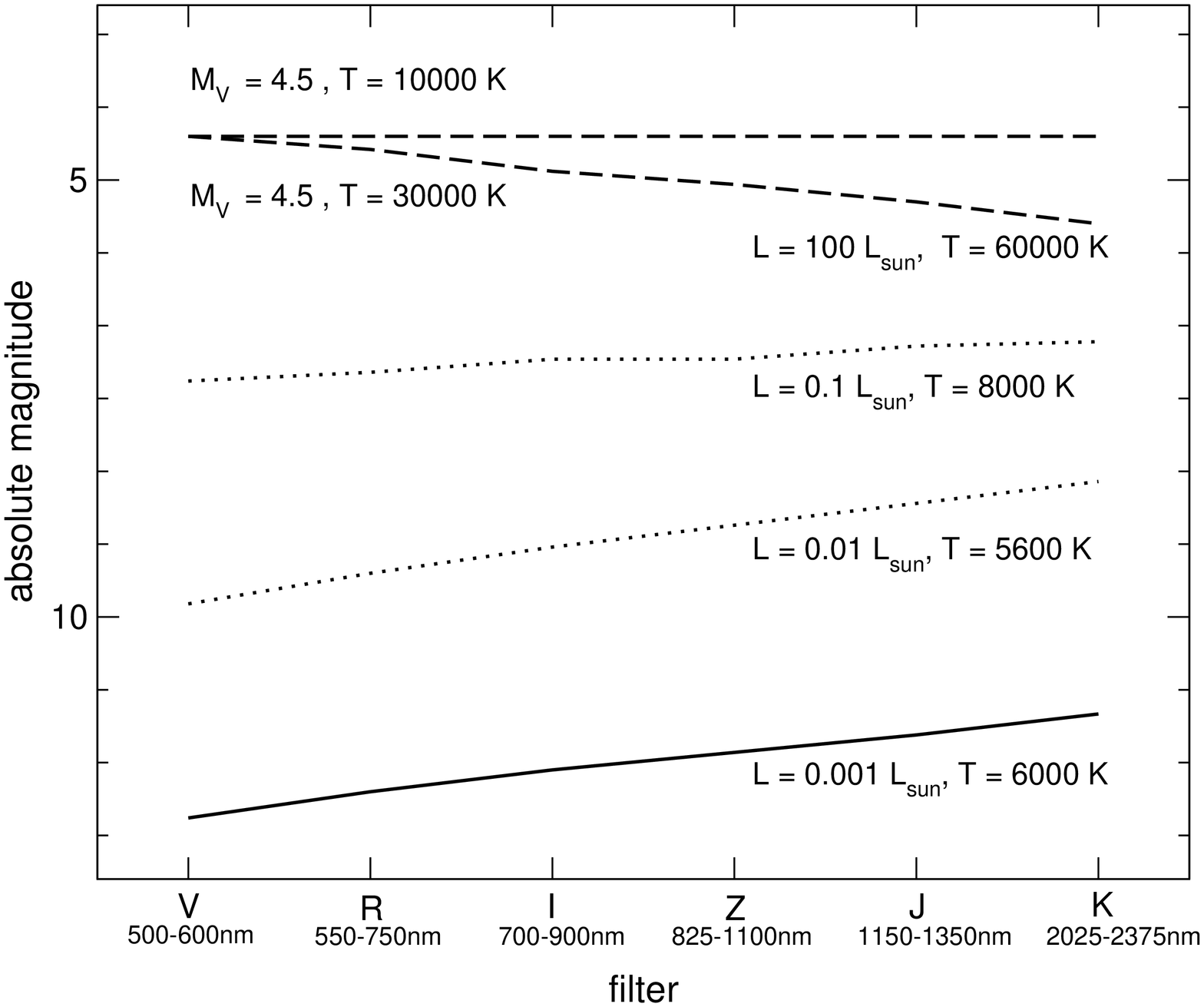}}
\quad
\resizebox{\columnwidth}{!}{\includegraphics[angle=-90]{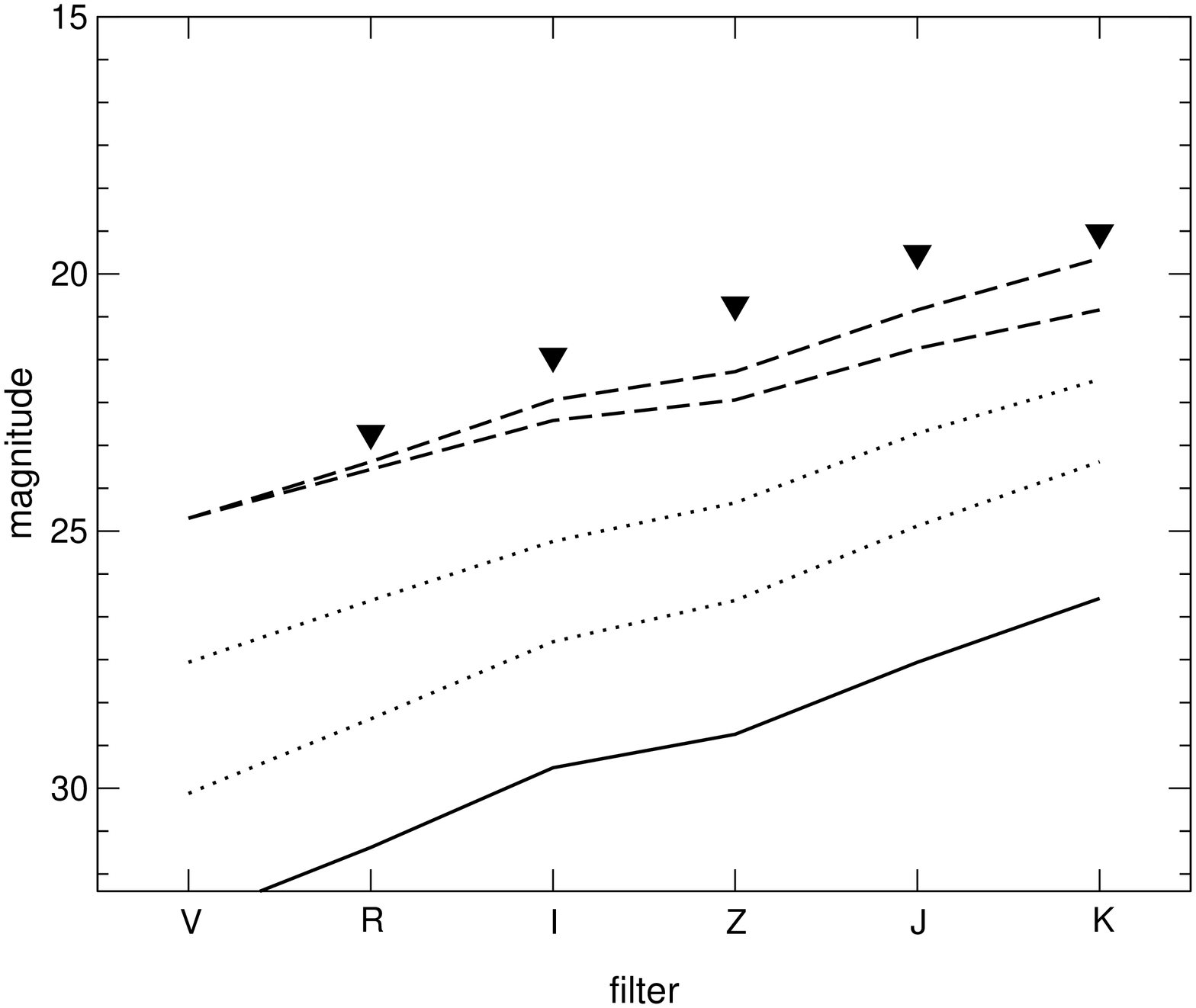}}

\caption[]{{\it Top panel:} Absolute magnitudes for the theoretical models for the
  donor stars in 40 minute periods ultra-compact X-ray binaries. Solid
  line: the hot white dwarf model from \scite{2002ApJ...577L..27B}.
  Dotted lines: the pre period--minimum models for XB~1832--330 of
  (left two triangles in the top panel of figure 16 of
  \pcite{2002ApJ...565.1107P}). Dashed lines: estimates for the
  outburst disc spectrum, based on an absolute visual magnitude of 4.5
  and temperatures of 10000 and 30000 K. The top dashed line also
  gives the absolute magnitudes of the donor in case it would be a
  luminous, hot helium star (from \pcite{1989SvA....33..606T}).  The
  passbands we used for the filters are indicated (we assumed perfect
  step function filters). {\it Bottom panel:} The estimated magnitudes
  for the same models as in Figure~\ref{fig:abs_mags}, assuming the
  system is close to the Galactic centre and taking the reddening into
  account ($A_{\rm V}$ = 5.6, based on the estimated ${\rm N_H}\sim
  1\times10^{22}$ cm$^{-2}$). The filled triangles represent the upper
  limits on the counterpart of XTE~J1751--305. }
\label{fig:abs_mags}
\end{figure}

To evaluate the constraints these upper limits can place on the
properties of the system, we consider the possible evolutionary states
for XTE~J1751--305. To arrive at an approximately forty minute orbital
period X--ray binary there are essentially three evolutionary paths.
The first starts from a detached white dwarf -- neutron star binary,
which is brought into contact by angular momentum loss due to
gravitational wave radiation.  For a discussion of this scenario for
ultra--compact X--ray binaries see \scite{2002A&A...388..546Y}. The
mass of the donor in this case would be $\sim$0.02 \msun\,and the mass
transfer rate $\sim$10$^{-11}$ \msun yr$^{-1}$. Recently,
\scite{2002ApJ...577L..27B} discussed XTE~J1751--305 and the similar
accreting millisecond X--ray pulsar XTE~J0929--314 and calculated
models for hot white dwarf donors, rather than the previously used
zero--temperature models. He finds a luminosity of the order of
10$^{-3}$ L$_{\odot}$ and an effective temperature of $\sim$6000 K for
the donors.

It is also possible to form a forty minute binary from a helium star
that transfers matter to a neutron star (e.g.
\pcite{1986AA...155...51S,1989SvA....33..606T}).  Such a scenario goes
through the forty minute period range twice. The first time, the
system goes through the orbital period of 40 minutes while the orbital
period decreases, then the donor is a luminous, hot helium star of
about 0.6 \msun. The mass transfer rate at that stage is of the order
of $10^{-8}$ \msun\, yr$^{-1}$.  After having reached a period minimum
of about 10 minutes, the system returns to longer periods with a dim,
semi--degenerate helium star donor, not unlike the low--mass white
dwarfs in the first scenario.  

The third evolutionary scenario producing ultra--compact binaries
involves a main sequence star close to core hydrogen exhaustion that
starts mass--transfer to a neutron star
(\pcite{1987SvAL...13..328T,2002ApJ...565.1107P}). Such a system
evolves to shorter orbital periods than the standard period--minimum
for hydrogen rich stars, since more and more helium enriched layers
are reached when peeling off the star. These systems go through a
period minimum, again of the order of 10 minutes, and form white dwarf
like, low--mass, low--luminosity donor stars.
\scite{2002ApJ...565.1107P} discuss possible models for XB~1832--330
which has an orbital period close to forty minutes as well, so these
models should also be applicable to XTE~J1751--305. The donors in these
models have masses and mass transfer rates between 0.026 and 0.14
\msun\,and 4.5 and 22$\times10^{-11}$ \msun\,yr$^{-1}$, respectively.


The X--ray observations already place some constraints on the system.
The mass--function for the companion (1.3$\times10^{-6}$ M$_\odot$;
\pcite{2002ApJ...575L..21M}) suggests that the companion is a low--mass
star. If the companion were to be a 0.6 \msun\,helium star, the
inclination should be less than 1.6$^\circ$. The two pre
period--minimum models of \scite{2002ApJ...565.1107P} with donor
masses of 0.14 and 0.094 \msun\,would imply inclinations less than
6$^\circ$ and 9$^\circ$, respectively. \scite{2002ApJ...575L..21M}
infer a mass transfer rate of 2.1$\times$10$^{-11}$ \msun\,yr$^{-1}$,
assuming a neutron star radius of 10 km, a distance of 10 kpc and a
recurrence time of 3.8 yr.  Even though there is quite some
uncertainly in the distance and recurrence time, the high mass
transfer rate as expected from a helium star donor seems to be
unlikely, even more so, because at such high rates the system would be
expected to be a persistent X--ray source
(\pcite{1997PASJ...49...75T}).

Next, we estimate the expected absolute magnitudes of the different
donor star and accretion disk models in the optical and near--infrared.
Since the source was still in outburst when we obtained our
observations it is likely that the accretion disk dominates in the
optical and near--infrared. We assumed an absolute visual magnitude of
the disk of 4.5 based on the observed absolute magnitude of
4U~1916--053 of 5.3, which has a similar orbital period, but allowing
for the higher X--ray luminosity of XTE~J1751--305 (assuming that the
disk luminosity scales with the square root of the X--ray luminosity
see \pcite{1995xrb..book...58V}), and the fact that 4U~1916--053 has a
high inclination reducing its absolute visual magnitude. Since these
ultra--compact systems have to have small accretion disks, they are
expected to be hot (see, \pcite{1995xrb..book...58V}). We estimated the
absolute magnitudes for disks modelling them as simple blackbodies of
10000 and 30000 K. The results are shown in Figure~\ref{fig:abs_mags}
(dashed lines in the {\it top panel}).

The absolute magnitudes of the possible donor stars in a forty minute
binary, again assuming simple blackbody spectral energy distributions,
are calculated for the hot white dwarf scenario proposed by
\scite{2002ApJ...577L..27B} and for the two pre period--minimum models
of \scite{2002ApJ...565.1107P}. We did not consider their post--period
minimum model because according to \scite{2002ApJ...577L..27B} the
companions in X--ray binaries will never cool down so much. Also, such
a donor star would be very similar to the hot white dwarf donor model.
For the hot helium star donor, we used a model having a luminosity of
100 L$_\odot$ and T = 60000 K (see \pcite{1989SvA....33..606T}).
Interestingly, the absolute magnitudes of this model fall on top of
the 30000 K disc model. The absolute magnitudes of all four models are
shown in Figure~\ref{fig:abs_mags} ({\it top panel}).

To compare the theoretical models with our upper limits, we assumed
XTE~J1751--305 is near the Galactic Centre, with a distance modulus of
14.65 magnitudes. Furthermore, we inferred an absorption ($A_{\rm V}$ =
5.6), based on the measured ${\rm N_H}$ = $10^{22}$ cm$^{-2}$
(\pcite{2002ApJMiller1751}) and calculated the absorption in the other
bands according to the relations found by \scite{1985ApJ...288..618R}.
The results are shown in Figure~\ref{fig:abs_mags} ({\it bottom panel})
where the single symbols denote the upper limits. Unfortunately, our
upper limits do not constrain any of the models. Further deep
near--infrared imaging could, however, start to rule out several
possible models for the counterpart of the accreting millisecond
X--ray pulsar XTE~J1751--305. According to the models presented above,
XTE~J0929--314, having an orbital period of $\sim$40 minutes, a
reddening in the V band of $\sim$0.65 magnitudes
(\pcite{1990ARA&A..28..215D}) can't have a distance much larger than
the lower limit of 6.5 kpc derived by \scite{2002ApJ...576L.137G}.
Otherwise it would not have been detected in outburst at V=18.8
(\pcite{2002IAUC.7889....1G}).

\section*{Acknowledgments} 
\noindent 
We would like to thank Mark Leising en Matt Wood for obtaining a B and
V band image with the SARA telescope, the Director of ESO for granting
the DDT time which made the ESO observations possible, and the referee
for constructive comments.

\end{document}